\documentclass[aps,prb,reprint,amsmath,amssymb,showpacs,superscriptaddress]{revtex4-1}

\usepackage{amsmath}
\usepackage{amsfonts}
\usepackage{amssymb}
\usepackage{mathtools}
\usepackage{pict2e}
\usepackage{hyperref}
\usepackage{graphicx,xcolor}
\usepackage{bbding}
\usepackage{bm}
\usepackage{multirow}
\usepackage{grffile}
\usepackage{comment}

\usepackage{color}

\newcommand{\bs}[1]{\boldsymbol{#1}}

\begin{document}

\title{Fluctuations and Noise Signatures of Driven Magnetic Skyrmions}

\author{Sebasti{\' a}n A. D{\' i}az}
\affiliation{Theoretical Division and Center for Nonlinear Studies, Los Alamos National Laboratory, Los Alamos, New Mexico 87545, USA}
\affiliation{Department of Physics, University of California, San Diego, La Jolla, California 92093, USA}

\author{C.J.O. Reichhardt}
\affiliation{Theoretical Division and Center for Nonlinear Studies, Los Alamos National Laboratory, Los Alamos, New Mexico 87545, USA}

\author{Daniel P. Arovas}
\affiliation{Department of Physics, University of California, San Diego, La Jolla, California 92093, USA}

\author{Avadh Saxena}
\affiliation{Theoretical Division and Center for Nonlinear Studies, Los Alamos National Laboratory, Los Alamos, New Mexico 87545, USA}

\author{C. Reichhardt}
\affiliation{Theoretical Division and Center for Nonlinear Studies, Los Alamos National Laboratory, Los Alamos, New Mexico 87545, USA}

\date{\today}

\begin{abstract}

Magnetic skyrmions are particle-like objects with topologically-protected stability
which can be set into motion with an applied current.
Using a particle-based model we simulate current-driven magnetic skyrmions interacting with random quenched disorder
and examine the skyrmion velocity fluctuations
parallel and perpendicular to the
direction of  motion as a function of increasing drive.
We show that the Magnus force contribution to 
skyrmion dynamics
combined with the random pinning produces an isotropic
effective shaking temperature.
As a result,
the skyrmions form a moving crystal
at large drives instead of the
moving smectic state observed
in systems with a negligible Magnus force where the effective
shaking temperature is anisotropic.
We demonstrate that spectral analysis of the velocity noise
fluctuations
can be used to
identify dynamical phase transitions
and to extract information about the different dynamic phases,
and show how the velocity noise fluctuations are correlated with
changes in the skyrmion Hall angle,
transport features, and skyrmion lattice structure.
\end{abstract}

\maketitle

\section{Introduction}

Skyrmions are particle-like objects
that emerge due to collective interactions of
atomic-scale magnetic degrees of freedom,
and they are stabilized by their topological properties
\cite{1,2,3}.
An increasing number of magnetic systems capable of supporting
skyrmions have been identified,
including recent work on materials in which the skyrmions
are stable at room temperature
\cite{4,5,6,7}.
Skyrmions can also be readily set into motion with an applied current
\cite{8,9,10,11,12,13,14,15,16}, and this property, along with their size scale, makes
them promising candidates as information carriers for applications in
high-density magnetic storage \cite{17,18,19,20} and novel logic devices \cite{21,22}.

Similar to other particle-like objects such as
vortices in type-II superconductors \cite{24,25,26,27},
electrons in Wigner crystals \cite{28,29,30}, and colloids moving on a rough
substrate \cite{31,32,33},
skyrmions
exhibit depinning and collective dynamics
when driven over quenched disorder \cite{23}.
All of these systems can
exhibit dynamical transitions between distinct types of motion
as a function of
the strength of the driving force.
Such dynamic phases include
a low drive pinned phase,
a plastic or disordered flow phase in which there is a  combination of
moving and pinned particles \cite{23,24,31,32,34,35}, moving liquid phases
in which all the particles are in motion
but the overall moving structure is disordered \cite{23,24},
and high driving dynamically ordered or quasi-ordered states
including moving smectic states and moving anisotropic
crystals \cite{23,24,26,27,29,33,36,37,38,39,40,41,42,43}.
These different phases can be identified by direct visualization \cite{39,41},
changes in the structure factor of the particle configurations \cite{26,37,38,40,42,43}
and features in the transport curves or
differential conductivity \cite{23,24,27,29,34,36,38,40}.
The dynamic states can also be characterized
by measuring changes
in the noise fluctuations of the moving particles as a function
of increasing drive \cite{23}. 
One of the best examples of
a system in which noise fluctuations can be used to
identify the dynamical states is flux motion in type-II
superconductors, where the voltage
noise generated at  fixed current or drive
is produced by the velocity fluctuations of the vortices.
Both experiments and simulations show that in the
plastic flow regime the noise fluctuations typically exhibit a large low frequency
component
with a $1/f^{\alpha}$
characteristic \cite{25,38,43} where $\alpha = 1.0$ to $2.0$,
while in the moving
liquid phase the noise is white with $\alpha=0$
\cite{25,38}.
At higher
drives when
a moving ordered state forms,
a narrow band noise signature appears
with
peaks at
frequencies that are correlated with the
periodicity of the moving structure \cite{38,41,43,44,45,46}.

Koshelev and Vinokur \cite{36} argued that the transition from the disordered flow state to the ordered
flow state at higher drives in vortex systems can be understood
by
representing the fluctuating forces exerted by the pinning sites on the moving vortices
as an effective shaking temperature $T_{sh}$.
As the driving force $F_d$ increases, $T_{sh}$ decreases according to
$T_{sh}\sim 1/F_d$
since the more rapidly moving vortices
have less time to respond to the pinning forces.
At high drives the pinning sites act only
as a weak perturbation of magnitude $T_{sh}$
on the rapidly moving vortices,  and the mutual repulsion between vortices
dominates over the pinning forces so that the vortices form an ordered lattice.
In contrast, at lower drives the
fluctuations created by the pinning sites
are stronger and $T_{sh}$ becomes large enough to melt the vortex lattice,
in
analogy to the melting of an equilibrium system under ordinary thermal fluctuations.
Further theoretical work \cite{47,48}
showed that
due to the direction of the external  driving force,
the shaking temperature from the
pinning-induced dynamical perturbations is anisotropic,
and thus the dynamically ordered state is also anisotropic,
taking the form
of a moving smectic
or a moving anisotropic crystal.
Such states have been observed in experiment \cite{39} and simulations \cite{37,38,40}.
Computational studies  have also directly shown that the dynamic fluctuations
in the different phases are anisotropic \cite{40}
and that the shaking temperature decreases with increasing drive \cite{49}.

Skyrmions share many features with vortices in 
type-II superconductors.
Both are particle-like objects with repulsive interactions,
both form triangular lattices in the absence of quenched disorder, and
both can be set into motion with an applied current.
Thus, analyzing the noise fluctuations of moving skyrmions should
provide a valuable method for characterizing the skyrmion dynamics.
One key difference between
vortices and skyrmions
is that skyrmion motion typically has a large
Magnus component, while vortex
systems are generally in the  overdamped limit where the Magnus term is weak or absent.
The Magnus term
generates a skyrmion velocity component
that is perpendicular to the external force,
and it is not known
what effect this has on the noise fluctuations.
Due to the Magnus term,  the skyrmions move at an
angle, called the skyrmion Hall angle $\theta_{Sk}$,
with respect to the driving force.
In the absence of pinning $\theta_{Sk}$ has a fixed 
intrinsic value of $\theta^{int}_{Sk} = \arctan(\alpha_m/\alpha_d)$
where $\alpha_m$ is the strength of
the Magnus term and $\alpha_d$ is the damping
coefficient.
Both simulations \cite{16,50,51,52,53,54} and experiments \cite{14,15}
have shown that when pinning is present, 
$\theta_{Sk}$ is strongly drive dependent,
taking the value $\theta_{Sk} \approx 0$
just above depinning and
increasing to
$\theta_{Sk} \approx \theta^{int}_{Sk}$
at high drives where the dynamic fluctuations induced by the pinning are reduced.
This drive dependence is a result of the side-jump motions the skyrmions experience
when they interact with pinning sites,
which diminishes at higher drives
\cite{50,51,52,55}.
Previous simulation work showed
that in the presence of pinning the skyrmions form
a pinned skyrmion glass that depins plastically
into a disordered flowing state followed by
a transition to an ordered flowing state at high drives \cite{51}.
Unlike the overdamped vortices, which form a moving smectic state,
the skyrmions form
a moving isotropic crystal,
suggesting that the dynamic fluctuations and  shaking temperature
experienced by the moving
skyrmions are different in nature from those in the vortex system.
Previous numerical studies indicate
that skyrmions show a broad band noise signature  near depinning
and narrow band noise at high drives \cite{53};
however, these studies were limited in scope and did not
include diffusive measures that could indicate
how the fluctuations correlate with the structure of the moving skyrmions.

Here we examine skyrmions moving over random quenched disorder and
measure the velocity fluctuations both parallel and perpendicular
to the direction of skyrmion motion
for varied intrinsic skyrmion Hall angles and drives.
We use a
particle-based model
\cite{16,50,51,52,53,55} in which the skyrmion dynamics is described by a modified
Thiele equation \cite{55,56}.
In the overdamped limit the Hall angle is zero with respect to the driving
direction,
the
velocity fluctuations are highly anisotropic at all drives, and the system
forms a moving smectic state at large drives as indicated by the presence of
two dominant peaks in the structure factor.
When the Magnus term is finite, we find that the velocity fluctuations
are anisotropic 
in the plastic flow regime
but become isotropic at the transition to a moving ordered phase
in which the skyrmions form a moving triangular crystal
with equal weight in the six
Bragg peaks of the structure factor.
The
skyrmion
Hall angle is zero just at depinning
within the plastic flow regime
and gradually increases with increasing drive before saturating to a
value close to $\theta_{Sk}^{int}$
at the transition to the moving crystal state.
We show that the velocity noise fluctuations
undergo a crossover from a broad band
noise signature in the plastic flow phase to
narrow band noise in the moving crystal. 
In general the skyrmion system
exhibits a richer variety of narrow band noise than the vortex system,
such as switching events
that are associated with small rotations in the moving lattice.
We also show that at small but finite intrinsic
skyrmion
Hall angles,
multiple dynamical transitions can occur
in the moving state,
including a transition from disordered flow to a moving smectic
followed by a transition to a  moving crystal.  These transitions
can be detected through
changes in the velocity noise signal and jumps
in the skyrmion Hall angle.

The paper is organized as follows.  In Section II we describe the system and
our simulation method.
Section III covers
the dynamic phases observed for
varied intrinsic
skyrmion
Hall angles, the behavior of the velocity
fluctuations parallel and perpendicular to 
the skyrmion motion, and the correlation of these fluctuations with
changes in transport curves, particle structure,
and
$\theta_{Sk}$.
In Section III A we show how the different dynamic phases
produce distinct structure factor
signatures, while in Section III B
we examine the diffusive behavior in the moving frame and find that the
shaking temperature in the dynamically ordered phase
is generally isotropic for skyrmions
and anisotropic in the overdamped limit.
We consider the limit of small intrinsic skyrmion Hall angles in Section III C
and show that
the system exhibits both a moving smectic and a moving crystal phase.
In Section IV we measure the velocity noise signals both
parallel and perpendicular to the direction
of skyrmion motion and show
that there is
a transition from broad band noise
in the plastic flow region to a narrow band noise signal in the moving crystal state.
In Section V we provide a summary of our results.

\section{Model and Simulation}

We consider a two-dimensional system
in which the skyrmions are modeled as point particles
obeying dynamics that are described 
by a modified Thiele's equation which includes quenched disorder and
skyrmion-skyrmion interactions \cite{55,56}. 
The particle-based model is applicable under conditions where
excitation of internal modes in individual skyrmions can be neglected and
where
the distance between skyrmions
is comparable to or larger than the size of an individual skyrmion.
We simulate $N = 480$ skyrmions moving in
a system of size $L_x \times L_y$ with $L_x=34.64$ and $L_y=36$
containing $N_p = 259$ randomly distributed pinning sites.
The equation of motion of skyrmion $i$ is given by
\begin{equation}
\alpha_d \bs{v}_i - \alpha_m \hat{\bs{z}} \times \bs{v}_i  = \bs{F}_i^{ss} + \bs{F}_i^{p} + \bs{F}^{D}.
\end{equation}
Here $\bs{v}_i=d\bs{r}_i/dt$ is the instantaneous velocity
and $\bs{r}_i$ is the position of skyrmion $i$, while $\alpha_d=1.0$ is the damping
coefficient for the dissipative viscous force.
The second term on the left hand side is the Magnus force
which originates from the topological charge of each skyrmion.
It produces no work and causes the skyrmions to move in the direction perpendicular to
the net force.
The skyrmion-skyrmion interaction force \cite{51,55} is given by
$\bs{F}_i^{ss} = \sum_{j = 1}^{N} K_1(r_{ij}) \hat{\bs{r}}_{ij}$,
where $r_{ij} = |\bs{r}_i - \bs{r}_j|$, $\hat{\bs{r}}_{ij} = (\bs{r}_i - \bs{r}_j)/r_{ij} $,
and $K_1$ is the 
modified Bessel function of the second kind.
The randomly distributed pinning sites are modeled as
finite range parabolic traps that produce a pinning force  described by
$\bs{F}_i^{p} = \sum_{k = 1}^{N_p} (F_p/r_p) (\bs{r}_i - \bs{r}_k^{(p)}) \Theta( r_p - |\bs{r}_i - \bs{r}_k^{(p)}| )$,
where $F_p=1.5$ 
is the maximum pinning force,
$r_p=0.35$ is the radius of the pinning sites,
$\bs{r}_k^{(p)}$ is the location of the $k$-th pinning site,
and $\Theta$ is the Heaviside step 
function. The external driving force, $\bs{F}^{D} = F_{D} \hat{\bs{x}}$, represents
the Lorentz force exerted on the emergent quantized magnetic flux
carried by each skyrmion by
an applied electric current \cite{3,9}.
The dynamics of superconducting vortices are also described by Eq. (1) in the
limit where the coefficient $\alpha_m$ of the Magnus term vanishes, since
the vortex dynamics are dominated by damping.  We refer to the case $\alpha_m=0$
as the overdamped vortex limit.

\begin{figure}[!h]
\centering
\includegraphics[width=\columnwidth]{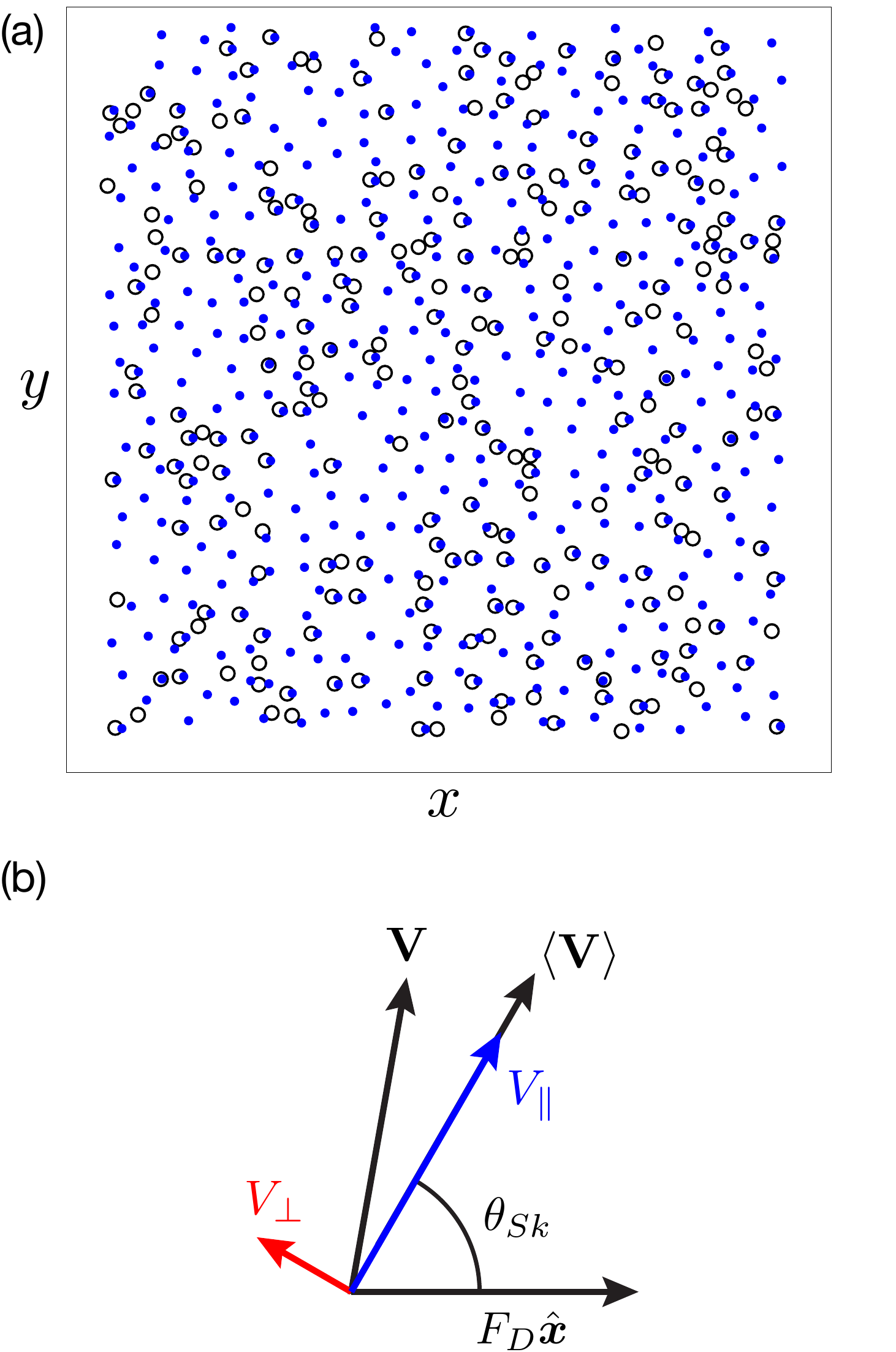}
\caption{
  (a) Real-space snapshot of the simulated system.
  Open black circles indicate pinning sites and filled blue dots
  indicate the instantaneous skyrmion positions.
  (b) Diagram illustrating the relationship between the driving force $F_D \hat{\bs{x}}$
  applied along the $x$ direction and the instantaneous skyrmion velocity ${\bf V}$
  averaged over all skyrmions.
  The skyrmion Hall angle $\theta_{Sk}$ varies with $F_D$, and the time-averaged
  skyrmion velocity $\langle {\bf V}\rangle$ is aligned with
  $\theta_{Sk}$.  The direction of ${\bf V}$ fluctuates around $\theta_{Sk}$ and we
  resolve ${\bf V}$ into its components $V_{||}$ and $V_{\perp}$ that are
  parallel and perpendicular, respectively, to the direction defined by
  $\theta_{Sk}$.
}
\label{fig:System}
\end{figure}

We measure ${\bf V}=N^{-1}\sum_{i=1}^N\bs{v}_i$,
the instantaneous velocity averaged over all of the skyrmions, as well as
its time-averaged value $\langle {\bf V}\rangle$.
In the absence of pinning sites,
the angle $\theta$ between ${\bf V}$ and the driving force is equal to
$\theta_{Sk}^{int}$, the intrinsic skyrmion Hall angle.
When pinning is present, however,
$\theta$ fluctuates around the drive-dependent value of $\theta_{Sk}$.
We resolve ${\bf V}$ into components
  $V_{||}={\bf V} \cdot {\hat{\bs x}}\cos{\theta_{Sk}}+{\bf V} \cdot {\hat{\bs y}}\sin{\theta_{Sk}}$ and
$V_{\perp}={\bf V} \cdot {\hat{\bs y}}\cos{\theta_{Sk}}-{\bf V} \cdot {\hat{\bs x}}\sin{\theta_{Sk}}$
that are
parallel and perpendicular, respectively, to the direction of $\theta_{Sk}$.
Figure~\ref{fig:System} shows an image of our system as well as a diagram illustrating
the relationship between ${\bf V}$, $V_{||}$, $V_{\perp}$, $\theta_{Sk}$, and the
driving force $F_D$ which is applied along the $x$ direction.

To investigate the effect of the Magnus term, we perform simulations
for different values of $\alpha_m$ corresponding
to intrinsic skyrmion Hall angles of $\theta^{int}_{Sk}=0^\circ$, $10^\circ$,
$20^\circ$, $30^\circ$, $45^\circ$, $60^\circ$, $70^\circ$, and $80^\circ$.
We initialize the skyrmions in a triangular lattice and slowly increase $F_D$
from $F_D=0$ to $F_D=8.0$ in increments of $\Delta F_D=0.01$
every
$5 \times 10^5$
simulation time steps.  
For each drive increment, we allow the system to equilibrate during $5 \times 10^4$
simulation time steps, compute $\theta_{Sk}$ for the equilibrated system based on the
direction of
$\langle{\bf V}\rangle$,
and then use
$\theta_{Sk}$ to measure $V_{||}$ and $V_{\perp}$
during the remaining
$4.5 \times 10^5$
simulation time steps.
From the resulting time series we construct
$\sigma_{||}$ and $\sigma_{\perp}$, the standard deviations of $V_{||}$ and $V_{\perp}$,
respectively.
We also obtain
$S_{||}(\omega)$ and $S_{\perp}(\omega)$,
the power spectral densities of the fluctuations
in the parallel and perpendicular velocity components.
Using the instantaneous positions of all the skyrmions,
we calculate the mean squared displacements  $\Delta_{||}$ and $\Delta_{\perp}$
parallel and
perpendicular to $\langle {\bf V}\rangle$, as well
as the static structure factor
$S(\bs{q}) = \left | N^{-1} \sum_{i = 1}^N \exp \left [ i \bs{q}\cdot\bs{r}_i \right ] \right |^2$.  We determine $P_n$, the fraction of $n$-fold coordinated skyrmions, according
  to $P_n=N^{-1}\sum_{i=1}^N \delta(n-z_i)$ where $z_i$ is the coordination number of
  skyrmion $i$ obtained from a Voronoi tessellation.

\section{Dynamic Phases}

\begin{figure}[!h]
\centering
\includegraphics[width=\columnwidth]{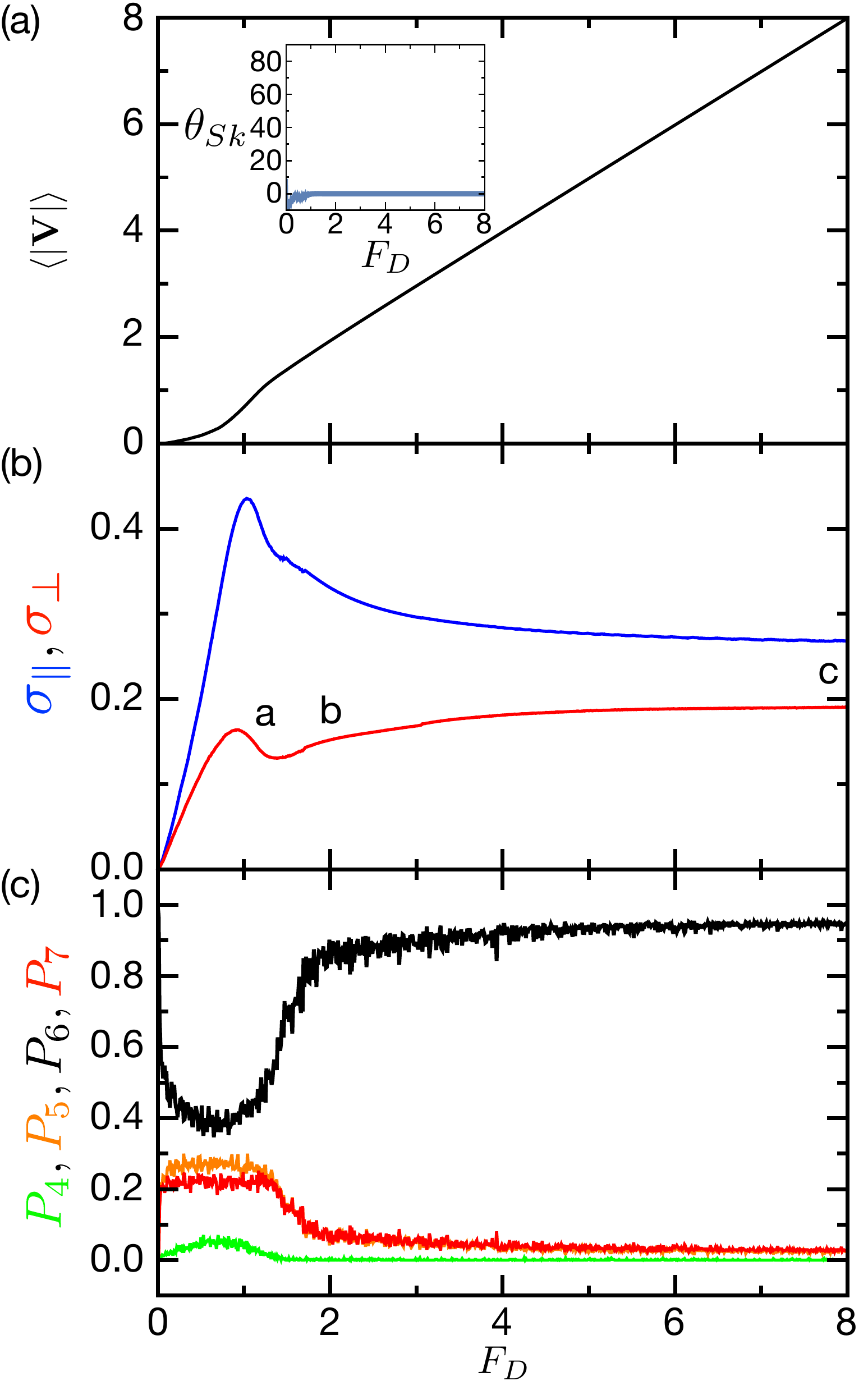}
\caption{
  Results from the overdamped limit $\alpha_m=0$ with $\theta^{int}_{Sk}=
  0^\circ$.
  (a) Magnitude of the time-averaged velocity
  $\langle |{\bf V}|\rangle$ vs
  driving force $F_{D}$. The inset shows $\theta_{Sk}$ vs $F_{D}$, where 
  there is no driving force dependence and $\theta_{Sk} = 0^\circ$.
  (b) Standard deviation $\sigma_{||}$ (upper blue curve) and
  $\sigma_{\perp}$ (lower red curve) of the parallel and
  perpendicular instantaneous velocity, respectively,
  vs $F_{D}$ showing a strong anisotropy.  The labels a, b, and c indicate the values of
  $F_D$ at which the structure factors in Fig.~\ref{fig:StructureFactors}(a,b,c)
  were obtained.
  (c) $P_4$ (lower green curve), $P_5$ (central light orange curve),
  $P_6$ (upper black curve), and $P_7$ (central dark red curve),
  the fraction of 4-, 5-, 6-, and 7-fold coordinated
  particles, respectively, vs $F_{D}$
  showing a transition to an ordered state near $F_{D} = 2.0$.
}
\label{fig:ThetaH_0}
\end{figure}

In Fig.~\ref{fig:ThetaH_0}(a) we plot
the magnitude of the time-averaged velocity
$\langle |{\bf V}|\rangle$ versus driving force $F_D$ for the overdamped
vortex limit with $\alpha_m=0$ and $\theta_{Sk}^{int}=0$.
For $F_D<2.0$, $\langle |{\bf V}|\rangle$ has a nonlinear shape
consistent with a plastic depinning process, and crosses over to a linear
dependence on driving force for $F_D \geq 2.0$.
As shown in the inset of Fig.~\ref{fig:ThetaH_0}(a),
$\theta_{Sk} = 0$ for all $F_{D}$,
indicating that the particles
are moving in the direction of the applied drive, as expected for the
vortex limit.
In Fig.~\ref{fig:ThetaH_0}(b), the plot of $\sigma_{||}$ and $\sigma_{\perp}$ versus $F_D$
indicates that the velocity fluctuations are strongly
anisotropic, with $\sigma_{||}>\sigma_{\perp}$.
Both $\sigma_{||}$ and $\sigma_{\perp}$ reach peak values
near $F_D=1.0$, indicating a maximum in the plasticity of the flow.
This is followed by a decrease in $\sigma_{||}$ to a saturation value of
$\sigma_{||}\approx 0.27$ at high drives,  and a dip in $\sigma_{\perp}$ followed
by a gradual increase to a saturation value of $\sigma_{\perp} \approx 0.19$ at
high drives.
The plots of $P_4$, $P_5$, $P_6$, and $P_7$ versus $F_D$ in
Fig~\ref{fig:ThetaH_0}(c) show
that for $0.0 < F_{D} <
1.2$,
$P_6 \approx 0.4$ and
the vortex positions are strongly disordered.
For $F_{D} >
1.2$,
$P_{6}$ increases to a value of $P_6 \approx 0.9$ indicating that
dynamical reordering of the vortices has occurred,
while $P_{5}$ and $P_{7}$ drop to $P_{5,7} \approx 0.05$ and track each other
as the defects form paired 5-7 gliding dislocations.

\begin{figure}[!h]
\centering
\includegraphics[width=\columnwidth]{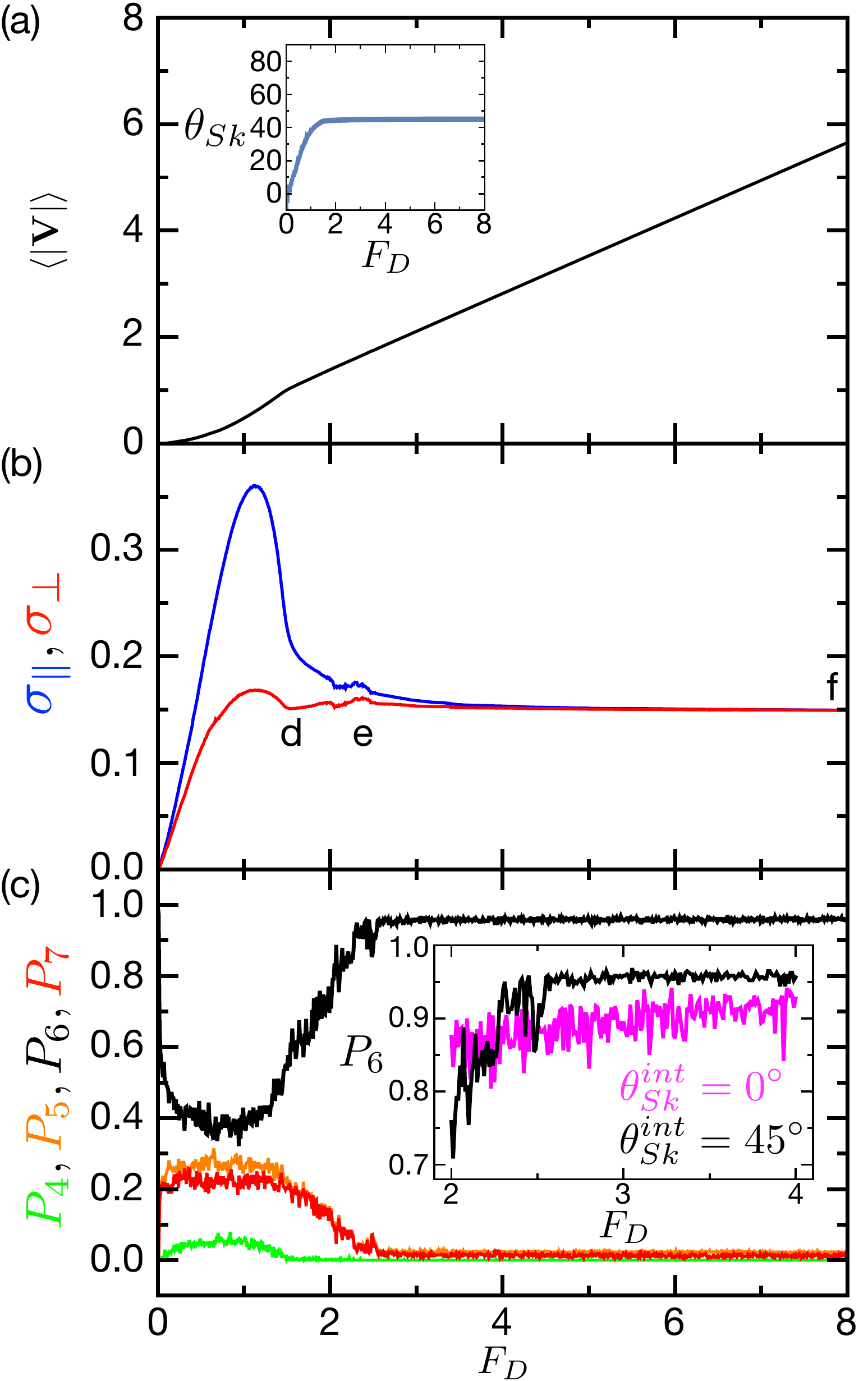}
\caption{
  Results from a system with $\theta^{int}_{Sk}=45^\circ$.
(a) $\langle |{\bf V}|\rangle$ vs $F_{D}$. The inset shows 
  $\theta_{Sk}$ vs $F_{D}$ where $\theta_{Sk}
  =0^\circ$
at $F_D=0$ and increases
  to $\theta_{Sk}=\theta_{Sk}^{int}$ when the system enters the moving crystal phase.
  (b)
  $\sigma_{||}$ (upper blue curve) and $\sigma_{\perp}$ (lower red curve)
  vs $F_{D}$.  $\sigma_{||} \approx \sigma_{\perp}$ 
  when the system enters the moving crystal state,
  where the velocity fluctuations are isotropic.
  The labels d, e, and f indicate the values of $F_D$ at which the structure factors
  in Fig.~\ref{fig:StructureFactors}(d,e,f) were obtained.
  (c)
  $P_4$ (lower green curve), $P_5$ (central light orange curve), $P_6$ (upper black
  curve), and $P_7$ (central dark red curve) vs $F_{D}$
  showing a transition to the moving crystal state near $F_{D} = 2.5$.
  Inset:
  $P_6$ vs $F_D$ for systems with $\theta^{int}_{Sk}=0^\circ$
  (gray curve) and $\theta^{int}_{Sk}=45^\circ$ (black curve)
  over the range $2.0 < F_D < 4.0$.  The $\theta_{Sk}^{int}=45^{\circ}$ system
  reaches a higher saturation value of $P_6=0.96$ at $F_D=2.5$.
} 
\label{fig:ThetaH_45}
\end{figure}

Figure~\ref{fig:ThetaH_45} illustrates the same quantities as above in a system with
a finite Magnus term where $\theta^{int}_{Sk}=45^{\circ}$.
In Fig.~\ref{fig:ThetaH_45}(a),
the shape of $\langle |{\bf V}|\rangle$ versus $F_D$ 
is nonlinear for $F_D<2.0$, indicating a plastic depinning process
similar to that of the $\theta^{int}_{Sk}=
0^\circ$
case shown
in Fig.~\ref{fig:ThetaH_0}(a), but
with a lower overall magnitude of $\langle |{\bf V}|\rangle$.
The plot of $\sigma_{||}$ and $\sigma_{\perp}$ versus $F_D$
in Fig.~\ref{fig:ThetaH_45}(b) shows that
$\sigma_{||} > \sigma_{\perp}$
only for $F_D<2.5$ in the disordered flow regime, while for $F_D \geq 2.5$ when
the system has dynamically ordered, $\sigma_{||} \approx \sigma_{\perp}$, indicating
that the velocity fluctuations are isotropic unlike in the vortex case.
In Fig.~\ref{fig:ThetaH_45}(c), we plot $P_4$, $P_5$, $P_6$, and $P_7$ versus
$F_D$.
The disordered plastic flow with low $P_6$ and high $\sigma_{||}$ persists out
to slightly higher drives in the $\theta_{Sk}^{int}=45^\circ$ system compared to
the vortex system, as indicated by the fact that the peak in $\sigma_{||}$ occurs
at $F_D=1.1$ in Fig.~\ref{fig:ThetaH_45}(b) but at $F_D=1.0$ in
Fig.~\ref{fig:ThetaH_0}(b).  The dynamic reordering also shifts to higher drives,
falling at $F_D=2.0$ in Fig.~\ref{fig:ThetaH_0}(c) but at $F_D=2.5$ in
Fig.~\ref{fig:ThetaH_45}(c); however, although $P_6$ continues to slowly increase
with $F_D$ in the vortex system above the dynamic reordering transition, in the
$\theta_{Sk}^{int}
=45^\circ$
  system $P_6$ reaches a higher saturation value
of $P_6=0.96$ at $F_D=2.5$ and does
not evolve as $F_D$ further increases,
as shown in the inset of Fig.~\ref{fig:ThetaH_45}(c).
This difference is due to the anisotropic velocity
fluctuations in the vortex case, where the
dislocations remaining after the dynamic reordering process has occurred undergo
gliding motion and can only slowly annihilate each other, and the
isotropic velocity fluctuations in the skyrmion case,
where the remaining dislocations can climb under effectively thermal excitations
and annihilate each other much more efficiently.
The inset of Fig.~\ref{fig:ThetaH_45}(a) shows that
$\theta_{Sk} \approx 0$ at small $F_D$
and gradually increases to $\theta_{Sk}=\theta^{int}_{Sk}=45^{\circ}$, reaching its
saturation value at $F_D=1.5$ where the difference between $\sigma_{||}$ and
$\sigma_{\perp}$ in Fig.~\ref{fig:ThetaH_45}(b) has become small.
This indicates that the direction of skyrmion motion in the plastic flow state
gradually rotates until the drive is high enough that all the skyrmions have
depinned.
The drive dependence of $\theta_{Sk}$ is in agreement with
previous simulation
studies \cite{51} and experiments \cite{14}
which show an increase in $\theta_{Sk}$ with driving force and a saturation
to the intrinsic value $\theta_{Sk}^{int}$ at high drives.

\begin{figure}[!h]
\centering
\includegraphics[width=\columnwidth]{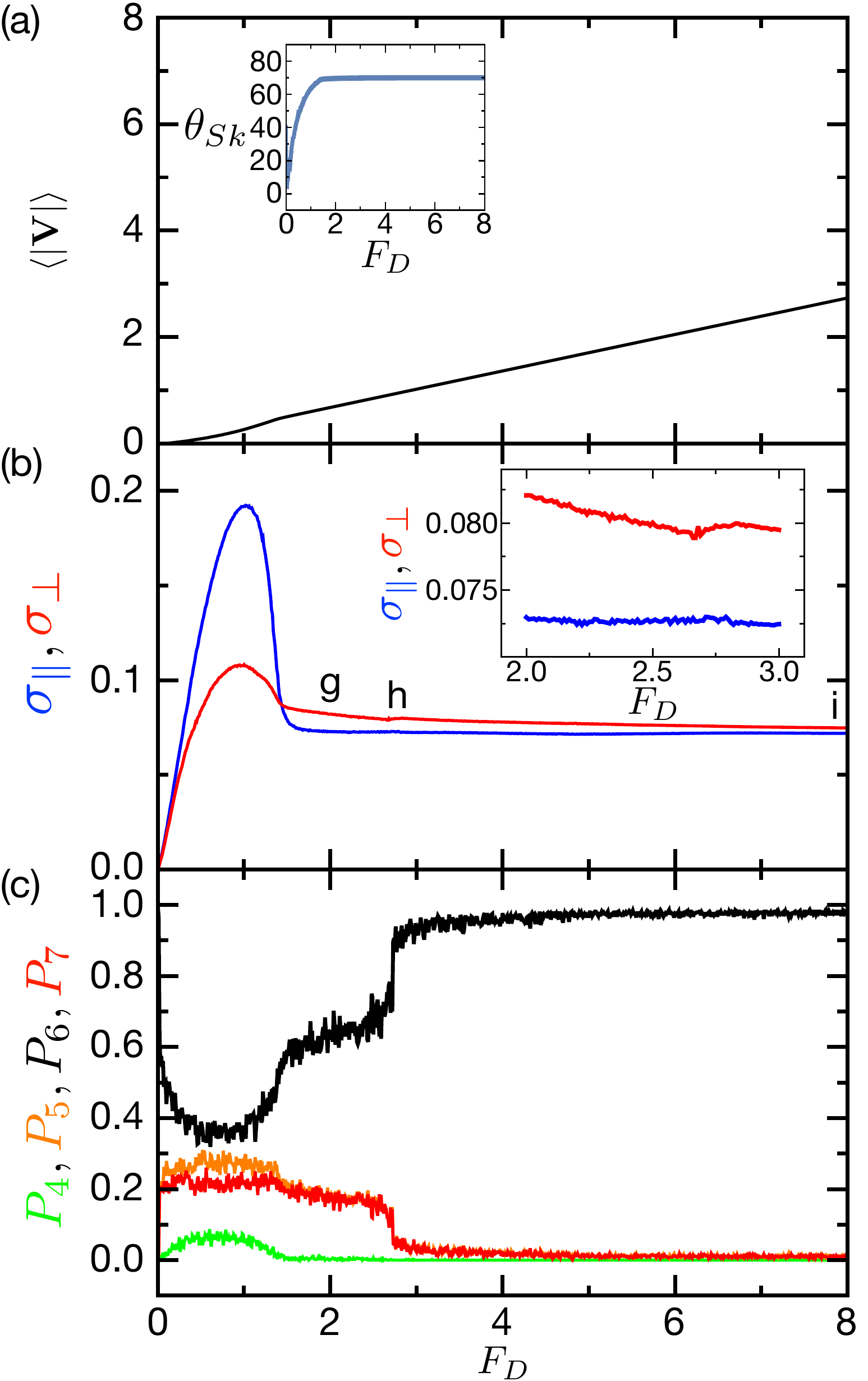}
\caption{
  Results from a system with $\theta_{Sk}^{int}=70^\circ$.
  (a) $\langle |{\bf V}|\rangle$
  vs $F_{D}$. The inset shows $\theta_{Sk}$ vs $F_D$ where $\theta_{Sk}=0$
  at $F_D=0$ and increases to
  $\theta_{Sk}=\theta_{Sk}^{int}$
  when the system enters
  the moving crystal phase.
  (b) $\sigma_{||}$ (upper left blue curve) and $\sigma_{\perp}$ (lower left red curve)
  vs $F_{D}$
  showing that the velocity fluctuations become isotropic at high drives.
  The labels g, h, and i indicate the values of $F_D$ at which the structure
  factors in Fig.~\ref{fig:StructureFactors}(g,h,i) were obtained.
  Inset: A zoom of the main panel highlighting the small jump in $\sigma_{\perp}$ at
  the transition from the moving liquid to the moving crystal.
  (c) $P_4$ (lower green curve), $P_5$ (central light orange curve), $P_6$ (upper
  black curve), and $P_7$ (central dark red curve) vs $F_D$
  showing a multi-step transition to the moving crystal state near $F_D=2.75$.
  }
\label{fig:ThetaH_70}
\end{figure}

The same quantities as above for a system with $\theta^{int}_{Sk}=70^{\circ}$ appear
in Fig.~\ref{fig:ThetaH_70}.
$\langle |{\bf V}|\rangle$ versus $F_D$ in Fig.~\ref{fig:ThetaH_70}(a) has a
similar nonlinear shape for $F_D<2$ as found for the other values of
$\theta^{int}_{Sk}$, and at higher drives the magnitude of
$\langle|{\bf V}|\rangle$ is further decreased compared to the vortex system.
The inset of Fig.~\ref{fig:ThetaH_70}(a) indicates that at low drives $\theta_{Sk}=0$,
while the skyrmion Hall angle
saturates to the value $\theta_{Sk}=\theta_{Sk}^{int}
=70^{\circ}$
at $F_D=1.5$.
In Fig.~\ref{fig:ThetaH_70}(b), the plot of $\sigma_{||}$ and $\sigma_{\perp}$ versus
$F_D$ shows that there is a sharp decrease in both quantities
when the system enters the moving liquid phase at $F_D=1.5$, and that at high
drives $\sigma_{\perp}$ is slightly larger than $\sigma_{||}$.
The plot of $P_4$, $P_5$, $P_6$, and $P_7$
versus $F_D$ in Fig.~\ref{fig:ThetaH_70}(c) indicates that there are
two distinct disordered flowing phases.
For $0.3 < F_{D} < 1.5$, the system exhibits plastic flow (PF) where not all
of the skyrmions are moving, 
$\sigma_{\parallel}>\sigma_{\perp}$, $P_6 = 0.4$, $P_4>0$, and $P_5>P_7$.
In contrast, for $1.5<F_D<2.5$ the system forms what we call a moving liquid (ML) phase
in which all of the skyrmions have depinned and have gained some partial
order, with $P_6 \approx 0.625$, $P_4=0$, and $P_5=P_7$ indicating that all the
remaining dislocations have formed pairs.
For $F_{D} > 2.5$ the system dynamically orders into a moving crystal
as shown by the jump in $P_{6}$ to $P_6\approx 0.98$ and the drop in $P_5$ and $P_7$
to $P_5 = P_7 \approx 0.02$.
At the transition from the moving liquid to the moving crystal
there is also a small jump in $\sigma_{\perp}$, as shown in the inset
of Fig.~\ref{fig:ThetaH_70}(b).

\begin{figure*}[!t]
\centering
\includegraphics[width=\textwidth]{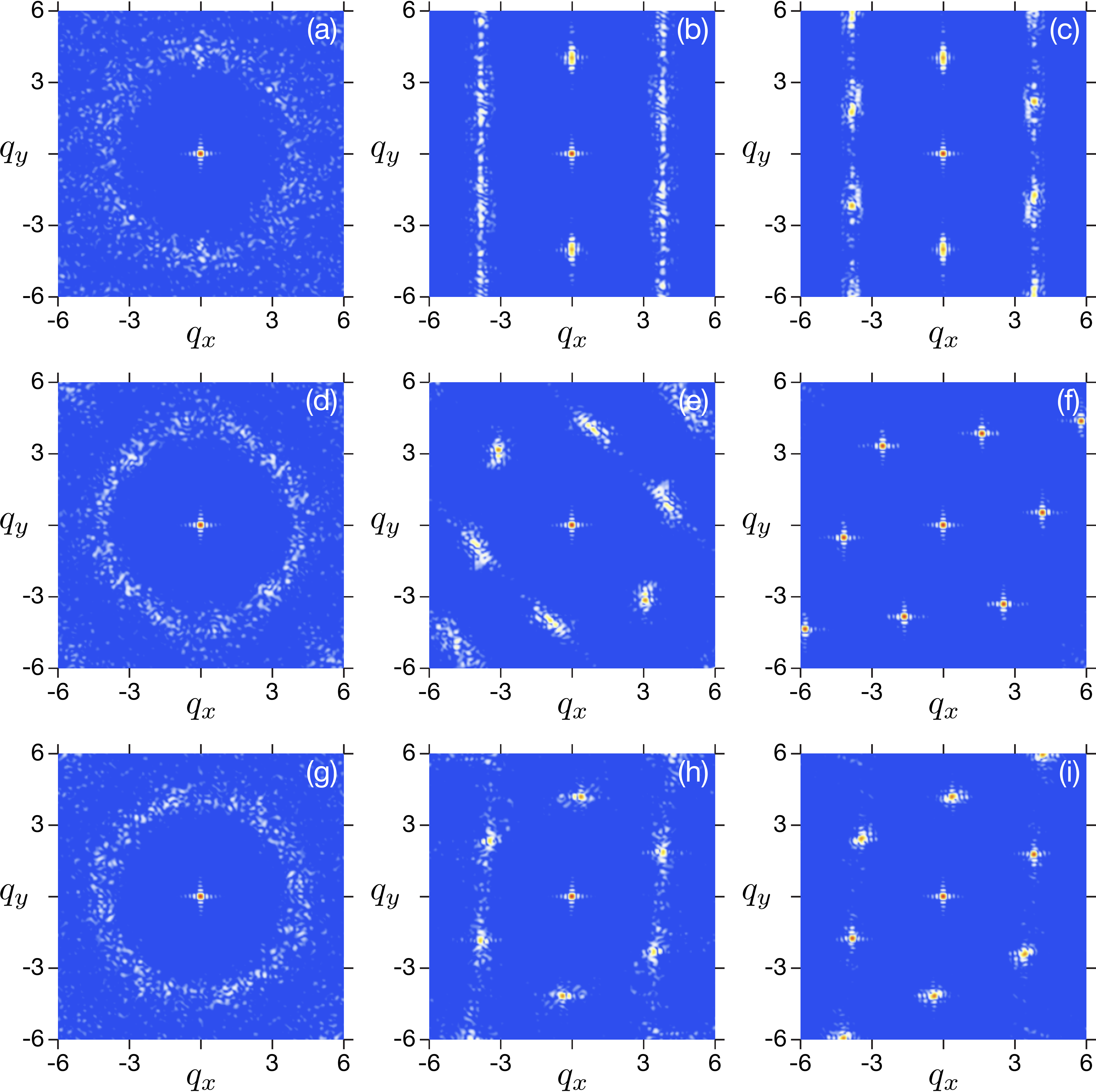}
\caption{
  Static structure factors $S(\bs{q})$.
  (a,b,c) The system in Fig.~\ref{fig:ThetaH_0} with $\theta_{Sk} = 0^\circ$
  at drives of (a) $F_D=1.3$ in the plastic flow state,
  (b) $F_D=2.0$ in the
  moving
  smectic state, and
  (c) $F_D=8.0$ in the moving anisotropic crystal state.
  (d,e,f) 
  The system in Fig.~\ref{fig:ThetaH_45} with $\theta_{Sk} = 45^\circ$
  at drives of (d) $F_D=1.6$ in the moving liquid state,
  (e) $F_D=2.4$ in a slightly anisotropic
  moving crystal state, and (f) $F_D=8.0$ in the moving crystal state.
  (g,h,i) The system in Fig.~\ref{fig:ThetaH_70} with $\theta_{Sk} = 70^\circ$
  at
  drives of (g) $F_D=2.0$ in the moving liquid state,
  (h) $F_D=2.8$ in a slightly anisotropic
  moving crystal state, and (i) $F_D=8.0$ in the moving crystal state.
}
\label{fig:StructureFactors}
\end{figure*}

\subsection{Moving Lattice Structure}

The different features in
the transport curves and velocity
noise fluctuations
correlate with
changes in
the static structure factor $S(\bs{q})$.
In Fig.~\ref{fig:StructureFactors}(a,b,c) we plot $S(\bs{q})$ for the overdamped
system with
$\theta^{int}_{Sk} = 0^{\circ}$
from Fig.~\ref{fig:ThetaH_0}
at drives of $F_{D} = 1.3$, $2.0$, and $8.0$, respectively.
At $F_{D}= 1.3$ the system is undergoing plastic flow and
$S(\bs{q})$
in Fig.~\ref{fig:StructureFactors}(a)
exhibits a ring feature indicative of an amorphous structure,
while at $F_{D} = 2.0$
in Fig.~\ref{fig:StructureFactors}(b) the system is in a
moving
smectic state in which
the particles travel in one-dimensional chains that slide
past one another, producing two peaks
in $S(\bs{q})$ along the $q_{x} = 0$ axis.
At $F_{D} = 8.0$ in Fig.~\ref{fig:StructureFactors}(c)
there is more order in the system as indicated by
the additional smeared peaks appearing in
$S(\bs{q})$;
however, the system is still in a moving smectic or a
strongly anisotropic moving crystal state
as
shown by the anisotropy in
$\sigma_{\perp}$ and $\sigma_{||}$ in Fig.~\ref{fig:ThetaH_0}(b).
This result is
in agreement with previous simulations and
experiments in the overdamped limit for vortex systems \cite{37,38,39}.

In Fig.~\ref{fig:StructureFactors}(d,e,f) we
plot $S(\bs{q})$ for the system
in Fig.~\ref{fig:ThetaH_45} with $\theta^{int}_{Sk} = 45^{\circ}$
at drives of $F_{D} = 1.6$, $2.4$ and $8.0$.
At $F_{D}  = 1.6$ in Fig.~\ref{fig:StructureFactors}(d)
the system is in a moving liquid phase,
while at $F_{D} = 2.4$
in Fig.~\ref{fig:StructureFactors}(e) the system
has dynamically ordered into a slightly
anisotropic moving crystal phase. Here $S(\bs{q})$ is rotated compared to
the
$\theta^{int}_{Sk} =
0^\circ$
case in Fig.~\ref{fig:StructureFactors}(b) since the skyrmion
lattice is moving at an angle with respect to the drive direction.
There is weak anisotropy in the moving structure as
indicated by the smearing of four of the
peaks in $S(\bs{q})$,
which is consistent with the small anisotropy
observed in $\sigma_{\perp}$ and $\sigma_{||}$ for this drive value
in 
Fig. ~\ref{fig:ThetaH_45}(b).
At $F_{D} = 8.0$
in Fig.~\ref{fig:StructureFactors}(f)
where $\sigma_{\perp} = \sigma_{\parallel}$,
the system forms a moving crystal state in which all six peaks in
$S(\bs{q})$ have almost equal weight.
In Fig.~\ref{fig:StructureFactors}(g,h,i) we plot
$S(\bs{q})$ for the system in Fig.~\ref{fig:ThetaH_70} with
$\theta^{int}_{Sk} = 70^\circ$ at drives of $F_{D} = 2.0$, $2.8$, and $8.0$, respectively.
Similar to the $\theta^{int}_{Sk}=45^\circ$ case,
at $F_D=2.0$ in Fig.~\ref{fig:StructureFactors}(g) the system forms
a disordered moving liquid state, at $F_D=2.8$ in Fig.~\ref{fig:StructureFactors}(h)
a slightly anisotropic moving crystal state appears, and
at high drives of $F_D=8.0$ in Fig.~\ref{fig:StructureFactors}(i),
an isotropic moving crystal forms
as indicated by the equal weight of the
six peaks in $S(\bs{q})$.

\begin{figure}[!h]
\centering
\includegraphics[width=0.9\columnwidth]{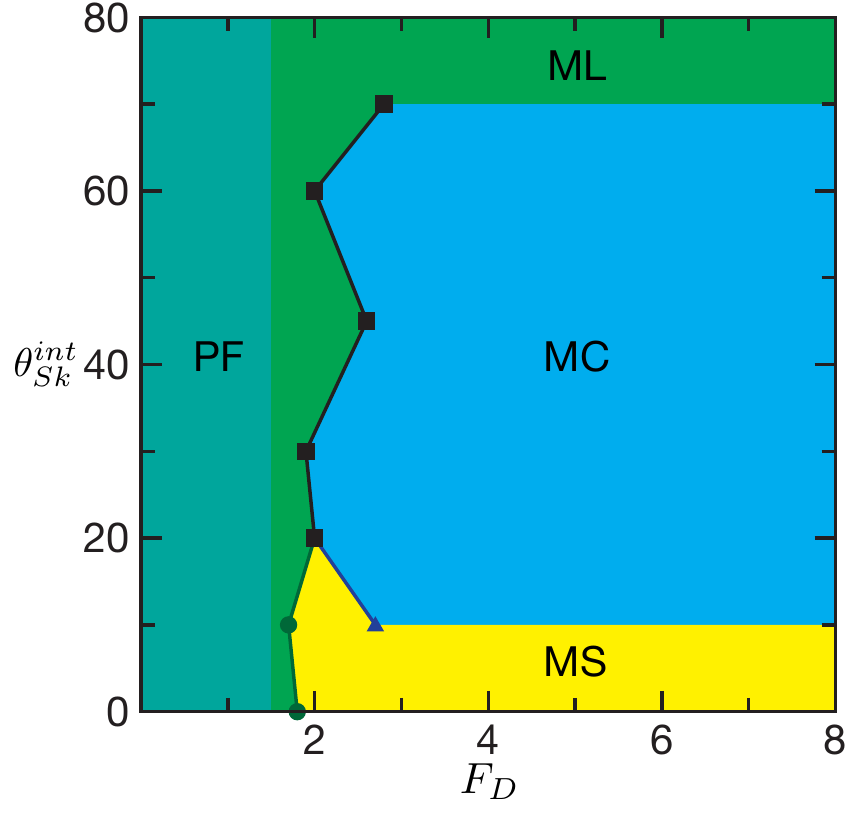}
\caption{
  Dynamic phase diagram as a function of the intrinsic
  skyrmion Hall angle $\theta^{int}_{Sk}$ vs
  the drive strength $F_D$. PF: plastic flow phase;
  ML: moving liquid phase; MS: moving smectic phase; and MC: moving crystal phase.
  The line separating PF from ML is at $F_D=F_p$.  For $F_D<F_p$, only a portion
  of the skyrmions are flowing, while for $F_D \geq F_p$, all of the skyrmions are
  moving.
  At high $\theta_{Sk}^{int}$ the MC phase is lost due to a Magnus melting effect.
}
\label{fig:DynamicPhaseDiagram}
\end{figure}

In Fig.~\ref{fig:DynamicPhaseDiagram}
we plot a dynamic phase diagram
as a function of $\theta^{int}_{Sk}$ versus $F_{D}$.
For $F_{D}/F_{p} < 1.0$ there is a mixture of pinned and moving
skyrmions and we find a plastic flow (PF) phase with a disordered structure.
In this
regime, the
velocity fluctuations are anisotropic with
$\sigma_{||} > \sigma_{\perp}$, and $\theta_{Sk}$
increases from zero as $F_{D}$ increases.
Just above $F_{D}/F_{p} = 1.0$, all the skyrmions
are moving but the structure is still disordered,
placing the system in a moving liquid (ML) phase.
The value of $\theta_{Sk}$ saturates to $\theta_{Sk}=\theta_{Sk}^{int}$ at the
onset of the ML phase.
For
$\theta^{int}_{Sk} < 10^\circ$,
at higher drives the system transitions into
the same moving smectic (MS) state found in the vortex system.
At
$\theta^{int}_{Sk} = 10^{\circ}$
we find that the ML-MS transition is followed at
slightly higher $F_D$ by a transition into a moving crystal (MC) phase.
In the MC phase the fluctuations are isotropic or only
slightly anisotropic with $\sigma_{\perp} \gtrsim \sigma_{||}$.
For $10^\circ < \theta^{int}_{Sk} \leq 70^\circ$
we find no MS phase and the system transitions directly from the ML phase to
the MC phase.
For $\theta^{int}_{Sk} = 80^{\circ}$, above the PF-ML transition
the system remains in the ML phase out to the highest values of $F_D$ we have
considered; however, in principle it is possible that the system could
dynamically order at much higher drives.
The loss of the MC phase at large values of
$\theta^{int}_{Sk}$ is due to the nondissipative nature of the
Magnus term which becomes increasingly
dominant as the intrinsic
skyrmion
Hall angle increases, creating larger dynamical
fluctuations that prevent the skyrmions from reordering
due to what we term a Magnus melting effect.

\subsection{Diffusion}

For vortex systems, the dynamic fluctuations of the vortices induced by 
interactions with the pinning sites are argued to act as an anisotropic
effective shaking temperature, with the largest fluctuations occurring along
the driving direction.
In order to test the nature of the fluctuations we
measure the mean squared displacement (MSD) of the skyrmions
by projecting the individual skyrmion displacements into the directions
parallel and perpendicular to the drive-dependent $\theta_{Sk}$ direction.
We define the parallel MSD in the center-of-mass frame of reference
at time $t$ as
$\Delta_\parallel(t) = N^{-1} \sum_{i = 1}^N \left[ \tilde{r}_{i,\parallel}(t) - \tilde{r}_{i,\parallel}(0) \right]^2$,
where 
$\tilde{r}_{i,\parallel}(t) = r_{i,\parallel}(t) - R^{\scriptscriptstyle\rm{CM}}_\parallel(t)$.
Similarly, the perpendicular MSD in the center-of-mass frame of reference is
given by
$\Delta_\perp(t) = N^{-1} \sum_{i = 1}^N \left[ \tilde{r}_{i,\perp}(t) - \tilde{r}_{i,\perp}(0) \right]^2$,
where
$\tilde{r}_{i,\perp}(t) = r_{i,\perp}(t) - R^{\scriptscriptstyle\rm{CM}}_\perp(t)$.
Here we have used the parallel and perpendicular components of the position
of the $i$-th skyrmion and of the center of mass $\bs{R}^{\scriptscriptstyle\rm{CM}}$. These components are given by $r_{i,\parallel}=\bs{r}_i \cdot {\hat{\bs x}}\cos{\theta_{Sk}}+\bs{r}_i \cdot {\hat{\bs y}}\sin{\theta_{Sk}}$, 
$r_{i,\perp}=\bs{r}_i \cdot {\hat{\bs y}}\cos{\theta_{Sk}}-\bs{r}_i \cdot {\hat{\bs x}}\sin{\theta_{Sk}}$, and similarly for $\bs{R}^{\scriptscriptstyle\rm{CM}}$. The position of the center of mass is $\bs{R}^{\scriptscriptstyle\rm{CM}} = N^{-1} \sum_{i = 1}^N \bs{r}_i$.

At long times, $\Delta_{||}(t)$ and $\Delta_{\perp}(t)$ exhibit
a power law behavior and are proportional
to $t^\alpha$, where $\alpha$ is the diffusion exponent.
We write $\Delta_{||}(t) \propto t^{\alpha_{||}}$ and $\Delta_{\perp}(t) \propto t^{\alpha_{\perp}}$.

\begin{figure*}[!t]
\centering
\includegraphics[width=\textwidth]{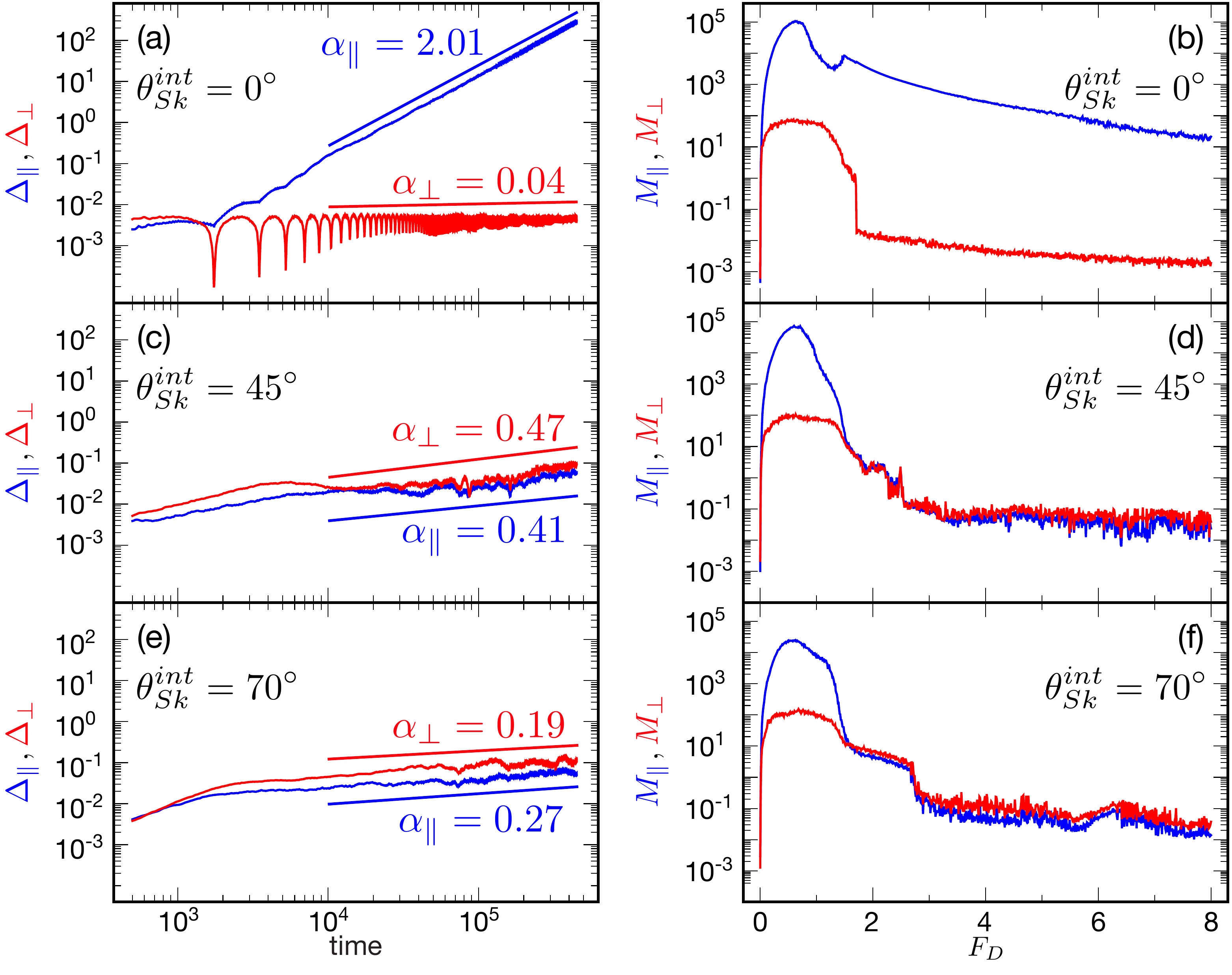}
\caption{
  (a,c,e) The parallel and perpendicular mean square displacements,
  $\Delta_{||}$ (blue) and $\Delta_{\perp}$ (red), vs time in simulation steps
  obtained at $F_D=4.0$.
  (a) At $\theta_{Sk}^{int}
  =0^\circ$
  in the moving smectic phase,
  the diffusion
  exponents are
  $\alpha_{||}=2.01$ and $\alpha_{\perp}=0.04$.
  (c) At $\theta_{Sk}^{int}=45^{\circ}$ in the moving crystal phase,
  $\alpha_{||}=0.41$ and $\alpha_{\perp}=0.47$.
  (e) At $\theta_{Sk}^{int}=70^{\circ}$ in the moving crystal phase,
  $\alpha_{||}=0.27$ and $\alpha_{\perp}=0.19$.
  (b,d,f) $M_{||}=\Delta_{||}(\tau)$ and $M_{\perp}=\Delta_{\perp}(\tau)$ vs $F_D$ with
  $\tau=5 \times 10^5$ simulation time steps.
  (b) $\theta_{Sk}^{int}=0^\circ$, (d) $\theta_{Sk}^{int}=45^\circ$, and
  (f) $\theta_{Sk}^{int}=70^\circ$.
  As the intrinsic skyrmion Hall angle increases, the Magnus term becomes
  dominant over the dissipation and the anisotropy in $\Delta_{||}$ and $\Delta_{\perp}$,
  which serve as measures of the effective temperatures parallel and perpendicular
  to $\theta_{Sk}$, is greatly reduced.
}
\label{fig:MSD}
\end{figure*} 

In Fig.~\ref{fig:MSD}(a)
we plot $\Delta_{||}$ and $\Delta_{\perp}$ versus time for the overdamped
case of $\theta^{int}_{Sk} = 0^{\circ}$ at a drive of $F_{D} = 4.0$ where the system
forms a moving smectic state.
Here $\Delta_{||} \gg \Delta_{\perp}$ and
power law fits of the long time behavior give
$\alpha_{||} \approx 2$, consistent with ballistic motion in the parallel direction,
and $\alpha_{\perp}=0.04$.
This result is in agreement with
previous simulation studies
measuring the diffusive behavior in the moving
smectic phase for vortices in type-II superconductors \cite{40}.
The superdiffusive behavior of $\Delta_{||}$ arises due to the fact that in the
moving
smectic state
the particles move in one-dimensional chains
that slip past each other due to their slightly different velocities.
As a result,
the distance between any two particles in different chains
in the moving center of mass frame
grows linearly with time.
Since there is no transverse hopping of particles from chain to chain,
diffusion perpendicular to the driving direction is strongly suppressed.
In Fig.~\ref{fig:MSD}(b) we plot $M_{||}$ and $M_{\perp}$, the value of
$\Delta_{||}$ and $\Delta_{\perp}$ at a time of $\tau=5 \times 10^5$ simulation
time steps,
versus $F_{D}$ for the $\theta_{Sk}^{int}=0^\circ$ system
in Fig.~\ref{fig:MSD}(a).
Here
$M_{||} > M_{\perp}$
for all $F_{D}$,
and there is a drop in $M_{\perp}$ at the ML-MS transition.
Within the ML phase,
$\Delta_{\perp}$ behaves diffusively while in the MS phase it has subdiffusive
behavior (not shown).

In Fig.~\ref{fig:MSD}(c) we plot $\Delta_{||}$
and $\Delta_{\perp}$ versus time for
a system with $\theta^{int}_{Sk} = 45^\circ$
at $F_{D} = 4.0$ in the moving crystal phase.
Here the displacements are isotropic and we find subdiffusive
behavior with $\alpha_{||} \approx \alpha_{\perp} \approx 0.45$.
In general, in the moving crystal phase
a small number of dislocations are present that can slowly climb or glide,
producing the weak subdiffusive behavior in both $\Delta_{||}$ and $\Delta_{\perp}$.
The plot of $M_{||}$ and $M_{\perp}$ versus $F_D$ in Fig.~\ref{fig:MSD}(d)
indicates that the displacements are anisotropic in the
PF phase for $F_{D} < 2.0$, and then become  isotropic in the MC  phase for $F_D \geq 2.0$.
These results indicate that
the
isotropic nature of the
effective shaking temperature in the moving skyrmion system
is responsible for the formation of an isotropic moving crystal phase.
Figure~\ref{fig:MSD}(e) shows $\Delta_{||}$ and $\Delta_{\perp}$ versus time
in the moving crystal phase at $F_D=4.0$
for a system with  $\theta^{int}_{Sk} = 70^{\circ}$.
We find
subdiffusive behavior in both directions with
$\alpha_{||} \approx \alpha_{\perp} \approx 0.22$, and
$\Delta_{\perp}$ is slightly larger than $\Delta_{||}$.
In Fig.~\ref{fig:MSD}(f) we show $M_{||}$ and $M_{\perp}$ versus $F_D$ for the same system,
where we observe a 
transition to isotropic diffusion at the higher drives.
At $\theta^{int}_{Sk} = 80^\circ$ where the moving liquid phase persists up to
high drives,
there is still a transition from anisotropic
diffusion in the plastic flow phase to isotropic
diffusion in the ML phase; however,
within the ML phase $\alpha_{||}$ and $\alpha_{\perp}$
have much higher values than what we observe in the MC phase for
smaller $\theta^{int}_{Sk}$.
This indicates that even within the
disordered flow regime,
the Magnus dominated dynamics modify the diffusive behavior
compared to what is observed in the overdamped case.

\begin{figure*}[!t]
\centering
\includegraphics[width=\textwidth]{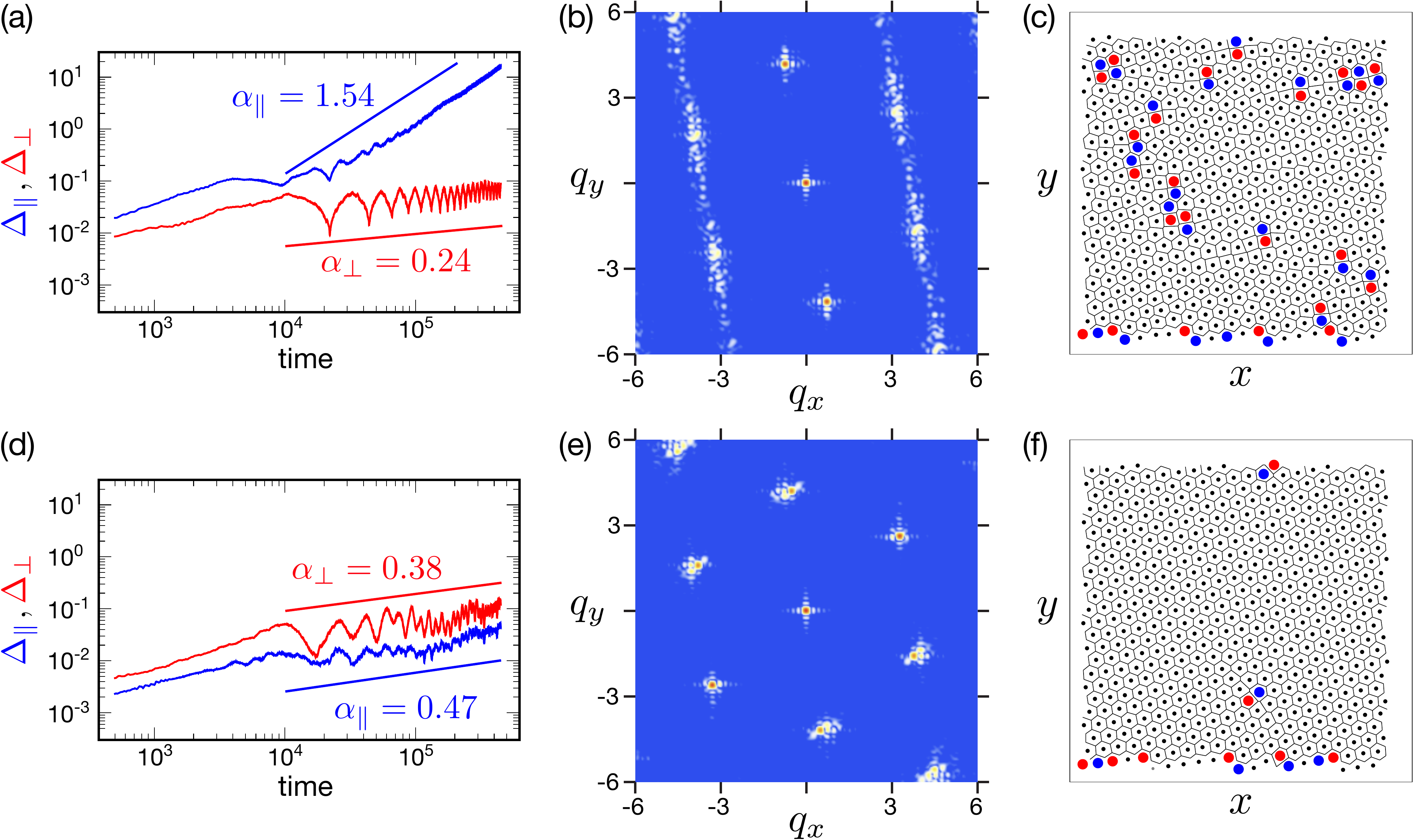}
\caption{
The transition from a moving smectic to a moving crystal for a  system with 
$\theta^{int}_{Sk} = 10^\circ$.
(a) $\Delta_{||}$ (upper blue line) and $\Delta_{\perp}$ (lower red line) vs time in simulation
time steps at $F_D=2.0$ in the moving smectic phase.  $\alpha_{||}=1.54$, indicating
superdiffusive behavior, while $\alpha_{\perp}=0.24$, indicating subdiffusive
behavior.
(b) The corresponding structure factor $S(\bs{q})$.
(c) Voronoi construction of the instantaneous skyrmion positions at $F_D=2.0$.
Black dots indicate sixfold-coordinated skyrmions, red dots indicate fivefold-coordinated
skyrmions, and blue dots indicate sevenfold-coordinated skyrmions.
In the smectic state, the defects combine into 5-7 pairs that glide along the
driving direction.
(d) $\Delta_{||}$ (lower blue line) and $\Delta_{\perp}$ (upper red line) vs time
in simulation time steps
at $F_{D} = 4.0$ in the moving crystal phase.
(e) The corresponding structure factor $S(\bs{q})$.  (f) The corresponding
Voronoi construction of the instantaneous skyrmion positions
at $F_D=4.0$ shows an almost completely ordered lattice.
}
\label{fig:Defects}
\end{figure*} 

\subsection{Small Intrinsic Skyrmion Hall Angles} 
 
The small
intrinsic
skyrmion Hall angle case of
$\theta^{int}_{Sk} = 10^{\circ}$ is particularly interesting since it exhibits
both a MS and a MC phase.
In Fig.~\ref{fig:Defects}(a) we plot
$\Delta_{\perp}$ and $\Delta_{||}$ as a function of time
for the
$\theta^{int}_{Sk} = 10^{\circ}$
system at $F_{D} = 2.0$ in the MS phase.
Here $\Delta_{||}$ is superdiffusive with $\alpha_{||}=1.54$ while
$\Delta_{\perp}$ is subdiffusive with $\alpha_{\perp}=0.24$.
Figure~\ref{fig:Defects}(b) shows that the structure factor at $F_D=2.0$ contains
prominent peaks indicative of a smectic ordering.
The moving smectic is tilted with respect to the $x$ axis
since the skyrmions are moving at
an angle $\theta_{Sk} \approx 9.8^{\circ}$ relative to the
driving direction.  This differs from the
MS that forms for $\theta^{int}_{Sk} = 0^{\circ}$,
shown in Fig.~\ref{fig:StructureFactors}(b),
which remains aligned with the drive direction.
In Fig.~\ref{fig:Defects}(c),
a Voronoi construction of the instantaneous skyrmion positions in the MS phase
indicates that the defects assemble into 5-7 pairs that glide along the
direction in which the skyrmions are moving.

At
$F_D=2.75$,
a transition occurs into
the MC state where the skyrmions
move at an angle
$\theta_{Sk} = 10^{\circ}$
to the driving direction,
indicating that the MS-MC transition
is also associated with a lattice rotation.
In Fig.~\ref{fig:Defects}(d) the plot of $\Delta_{\perp}$ and $\Delta_{||}$
versus time in simulation steps
at $F_D=4.0$ in the MC state shows that
the system is much more isotropic and that both $\Delta_{\perp}$ and $\Delta_{||}$
exhibit subdiffusive behavior with $\alpha_{||} \approx \alpha_{\perp} \approx 0.425$.
The  corresponding structure factor in Fig.~\ref{fig:Defects}(e) contains six peaks 
with equal weight.
In general the moving crystal phases are considerably more ordered than the
moving smectic phase, as illustrated by the Voronoi construction of the
instantaneous skyrmion positions in the MC phase shown in
Fig.~\ref{fig:Defects}(f), where there are 
significantly fewer 5-7 dislocation pairs compared to the MS phase.
Additionally, the dislocations
in the MC are not aligned with the direction of motion and can move
slowly in any of the six symmetry directions of the skyrmion lattice,
leading to the subdiffusive behavior.
It is possible that for even smaller but finite values of $\theta^{int}_{Sk}$,
MS-MC transitions could occur that are pushed to higher values of $F_D$ as
$\theta^{int}_{Sk}$ decreases.
Although skyrmion systems generally have $\theta^{int}_{Sk} > 25^{\circ}$,
a finite but small
Magnus term can still arise in some vortex systems,
so it may be possible to observe MS-MC transitions in superconducting systems,
particularly for weak pinning where large vortex velocities could be realized.

\section{Noise Signatures}

\begin{figure*}[!t]
\centering
\includegraphics[width=\textwidth]{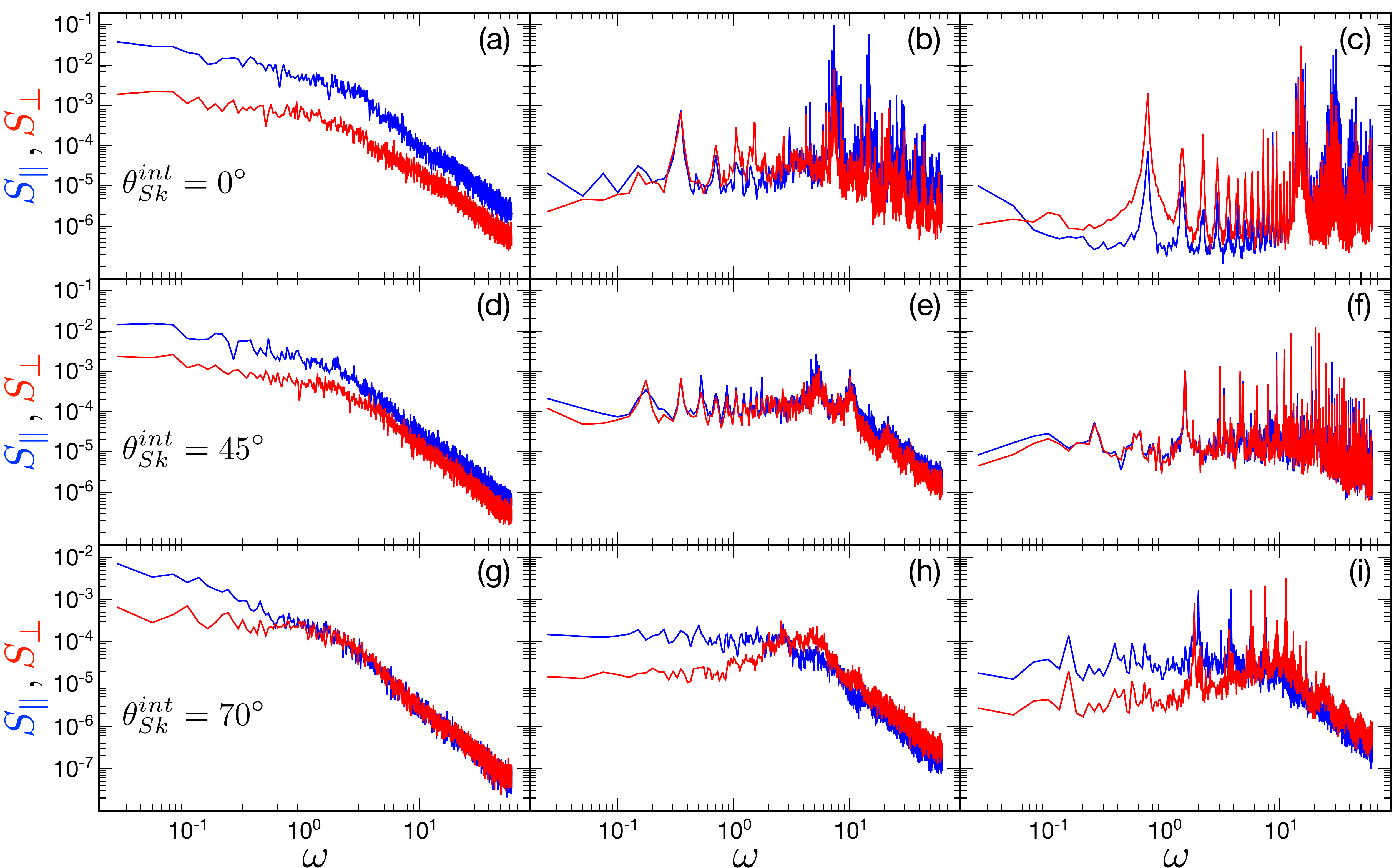}
\caption{
  Noise power spectral density plots $S_{||}(\omega)$ (blue) for velocity fluctuations
  parallel to $\theta_{Sk}$ and $S_{\perp}(\omega)$ (red) for velocity fluctuations
  perpendicular to $\theta_{Sk}$. 
  (a,b,c) A sample with $\theta_{Sk}^{int}=0^{\circ}$ at
  (a) $F_{D} =  1.0$ in the disordered flow state,
  (b) $F_{D} =  2.0$ in the moving smectic phase, and
  (c) $F_{D} =  4.0$ in moving smectic phase.
  (d,e,f) A sample with $\theta_{Sk}^{int}=45^{\circ}$ at
  (d) $F_{D} =  1.0$ in the disordered flow state,
  (e) $F_{D} =  2.0$ in the moving crystal phase, and
  (f) $F_{D} =  4.0$ in moving crystal phase.
  (g,h,i) A sample with $\theta_{Sk}^{int}=70^{\circ}$ at
  (g) $F_{D} =  1.0$ in the disordered flow state,
  (h) $F_{D} =  2.0$ in the moving
  liquid
  phase, and
  (i) $F_{D} =  4.0$ in the moving crystal phase.
}
\label{fig:PSD}
\end{figure*}

We next analyze
the velocity noise power spectral densities, $S_{||}(\omega)$ and $S_{\perp}(\omega)$,
computed from time series
taken at fixed $F_D$ of the velocity fluctuations parallel and perpendicular
to $\theta_{Sk}$, respectively.
In Fig.~\ref{fig:PSD}(a,b,c) we plot $S_{||}(\omega)$ and $S_{\perp}(\omega)$
for the overdamped case of
$\theta^{int}_{Sk} = 0^\circ$
at $F_{D} = 1.0$ in the plastic flow regime,
$F_{D} = 2.0$ in the moving smectic phase,
and at $F_{D} = 4.0$ in the moving smectic phase.
At $F_{D} = 1.0$, we find broad band noise in both directions with the
largest noise power in the parallel direction.
$S_{||}(\omega)$ 
has a $1/\omega$ feature at low frequencies
along with a $1/\omega^2$ tail at high frequencies, while
$S_{\perp}(\omega)$ becomes white at low frequencies.
At $F_{D} = 2.0$ in the moving smectic state, characteristic frequencies emerge
which become more pronounced at $F_D=4.0$.
We find a washboard frequency
of $\omega \approx 14.5$ at $F_D=4.0$
that is produced by the
perturbations of the periodic vortex lattice by the underlying pinning sites.
This frequency corresponds to
$\omega = 2\pi \langle |{\bf V}| \rangle/a$, where
$a$ is the characteristic spacing between the vortices in the driving direction.
We also observe 
a time-of-flight signal at $\omega=2\pi \langle |{\bf V}| \rangle/L_x$
which produces weaker peaks with $\omega<1$.
In superconducting systems both of these frequencies have been measured in
experiments and simulations.
In Fig.~\ref{fig:PSD}(d,e,f) we plot $S_{||}(\omega)$ and $S_{\perp}(\omega)$
for a system with
$\theta^{int}_{Sk} = 45^{\circ}$ at $F_{D} = 1.0$, $2.0$, and $4.0$, respectively.
At $F_{D} = 1.0$ in the plastic flow regime
we again find a broad band noise signal, but the anisotropy between
$S_{||}(\omega)$ and $S_{\perp}(\omega)$ 
is smaller than in the overdamped case.
In the moving crystal phase, at $F_{D} = 2.0$ the noise is isotropic
and a narrow band frequency emerges which becomes sharper at $F_{D} = 4.0$.
In Fig.~\ref{fig:PSD}(g,h,i) we show $S_{||}(\omega)$ and $S_{\perp}(\omega)$
for a system with $\theta^{int}_{Sk} = 70^{\circ}$
at $F_D=1.0$, 2.0, and 4.0, respectively.
At $F_{D} = 2.0$ the system is in a ML phase and the noise
power is flat at low frequencies,
while at $F_{D} = 4.0$ we find a strong washboard signal in both directions
with slightly higher signal strength at higher frequencies in the perpendicular
direction.
These results indicate that moving skyrmion lattices
can exhibit a narrow band noise signature
similar to that found in the vortex system,
and that analysis of this noise could be used to extract the lattice
constant and skyrmion velocity.

\begin{figure*}[!t]
\centering
\includegraphics[width=\textwidth]{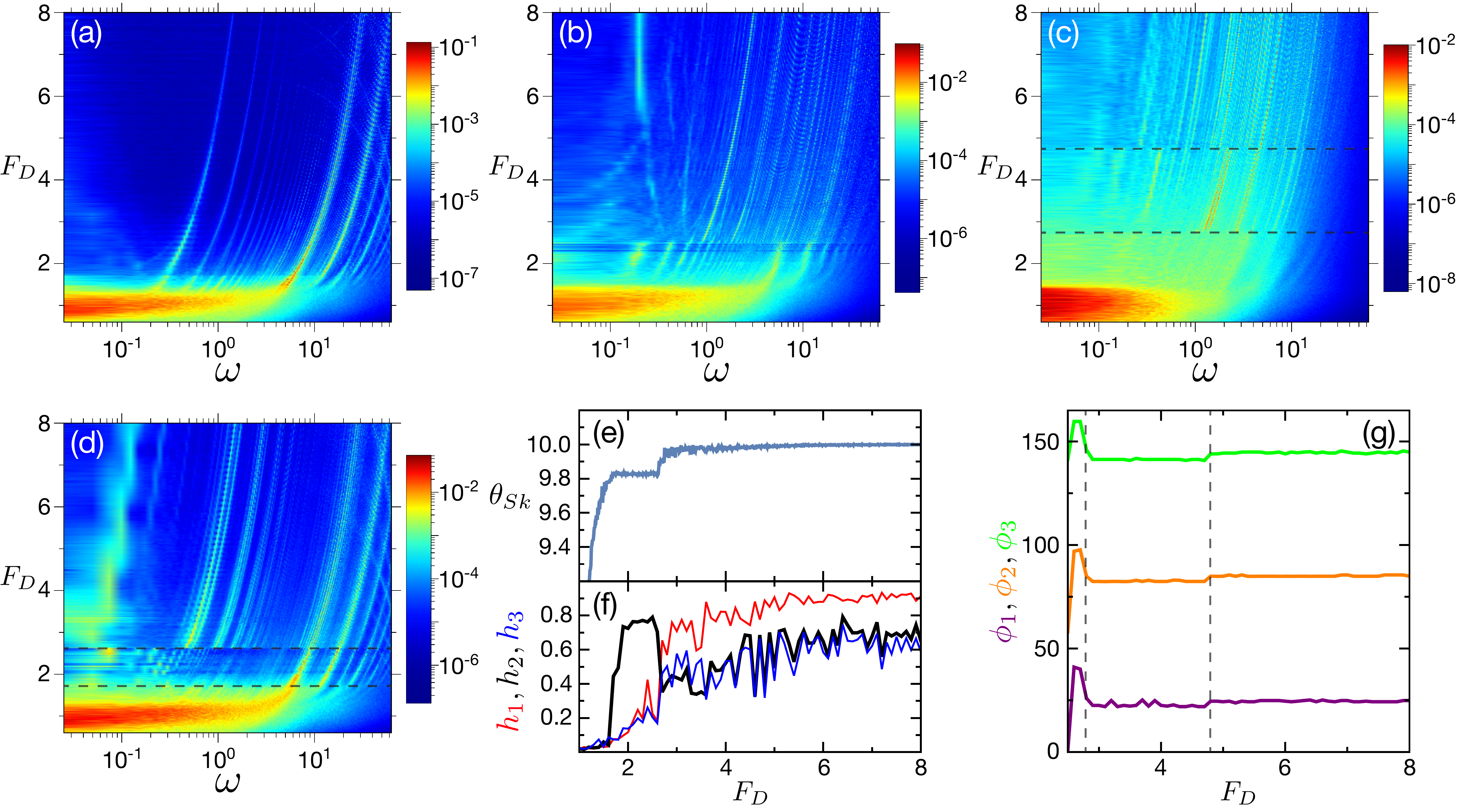}
\caption{
  (a,b,c,d) Heightfield plots of the power spectral density $S_{||}$ as a function
  of $F_D$ and $\omega$.
  (a) $\theta_{Sk}^{int} = 0^\circ$.
  (b) $\theta_{Sk}^{int} = 45^\circ$.
  (c) $\theta_{Sk}^{int} = 70^\circ$, where the two dashed lines
  indicate the switching events illustrated in panel (g).
  (d) $\theta_{Sk}^{int} = 10^\circ$, where
  there is a MS phase in the region between the two dashed lines.  
(e) $\theta_{Sk}$ vs $F_{D}$ at $\theta^{int}_{Sk} = 10^{\circ}$.
  (f) The structure factor peak heights
  $h_1$ (red), $h_2$ (black), and $h_3$ (blue)
  vs $F_{D}$ in a system with $\theta_{Sk}^{int} = 10^\circ$. Pairs of peaks $180^\circ$ apart are indistinguishable in the plotted range, shown are those corresponding to the first two quadrants.
  At large drives, peak $h_1$
  is at an angle $\phi_1 = 40^\circ$ with
  respect to the $x$ axis,
  peak $h_2$ is at $\phi_2 = 100^\circ$,
  and peak $h_3$ is at $\phi_3 = 160^\circ$.
  (g) The angles $\phi_1$, $\phi_2$, and $\phi_3$ at which
  three of the structure factor
  peaks appear vs $F_D$ 
  for the system in panel (c) with  $\theta_{Sk}^{int} = 70^\circ$.  The
two dashed lines indicate  switching events that are associated with a lattice rotation.
}
\label{fig:Noise-MS}
\end{figure*}

Another feature produced in the noise signatures by a finite Magnus term is
switching events that are associated with global rotations
in the moving skyrmion structures.
These switching events occur due to the drive dependence of the flow direction
generated by the Magnus term.
Such switching events do not occur
in an overdamped system since the lattice remains locked to the
driving direction once it has entered the MS phase.
In Fig.~\ref{fig:Noise-MS}(a) we plot a heightfield showing
the magnitude of the $S_{||}$ noise component
as a function of $F_{D}$ versus $\omega$ in the overdamped case
of $\theta^{int}_{Sk}=0^{\circ}$.
The plastic flow regime for $F_D<1.75$ is distinguished by large amplitude low
frequency noise,
while the onset of the narrow band noise at $F_D=1.75$ is indicated by a series of lines
that shift to higher frequencies with increasing $F_{D}$.
For $\theta^{int}_{Sk} = 45^\circ$, the heightfield plot of $S_{||}$ in
Fig.~\ref{fig:Noise-MS}(b)
shows several switching events that occur near $F_{D} = 2.25$
as indicated by the horizontal features.
Figure~\ref{fig:Noise-MS}(c) shows a heightfield plot of $S_{||}$ 
for $\theta^{int}_{Sk} = 70^\circ$.
Here the plastic phase appears as a strong signal at low frequency 
for $F_{D} < 1.5$,
while for $1.5 < F_{D} < 2.8$ the system is in the moving liquid phase.
The dashed line at $F_{D} = 2.85$ denotes the transition into the MC phase,
while the dashed line at $F_{D} = 4.85$ indicates a switching event.
We measure $\phi_n$, the angle between the $n$th structure factor peak and
the $x$ axis, for the $\theta^{int}_{Sk}=70^\circ$ system, and plot $\phi_1$, $\phi_2$,
and $\phi_3$ for the peaks in the first two quadrants versus $F_D$ 
in Fig.~\ref{fig:Noise-MS}(g).
The two dashed lines are at the same values
of $F_{D}$ as the dashed lines in Fig.~\ref{fig:Noise-MS}(c).
The first dashed line at $F_{D} = 2.85$ indicates
that just after the moving crystal forms,
there is a switching event in the lattice orientation, while
the dashed line at
$F_{D}  = 4.85$
shows
that there is a second switching event associated with a lattice rotation.
In Fig.~\ref{fig:Noise-MS}(c), several of the narrow band noise
peaks disappear above the second switching event.

Figure~\ref{fig:Noise-MS}(d) shows a heightfield plot of $S_{||}(\omega)$
versus $F_D$
at $\theta^{int}_{Sk} = 10^{\circ}$ where           
a frequency shift is associated with the MS-MC transition.
The region in which the MS phase appears is bounded by the
two dashed lines. Once the system
enters the MC phase, several additional narrow band noise peaks arise.
The plot of
$\theta_{Sk}$ versus $F_{D}$ for the $\theta^{int}_{Sk}=10^{\circ}$ system
in Fig.~\ref{fig:Noise-MS}(e)
shows that there is a jump in $\theta_{Sk}$ to $\theta_{Sk}=9.8^{\circ}$
at the transition to the MC phase.
We measure the magnitude $h_n$ of individual peaks in the structure factor
$S(\bs q)$
and plot $h_1$, $h_2$, and $h_3$ versus $F_D$ for the three peaks in the first and
second quadrants in Fig.~\ref{fig:Noise-MS}(f) for the same system.
In the MS phase, only
$h_2$
is large, indicating the strong asymmetry of the
weight of the peaks due to the smectic ordering.
At the transition
to the MC phase,
all the peaks in the structure factor have similar weights. 
These results indicate that switching events
in the noise spectral power measurements
can be used to deduce information about the lattice orientation.
 
\section{Summary}

We have numerically examined the velocity fluctuations parallel and
perpendicular to the direction of motion for
skyrmions driven over random disorder for different values
of the Magnus term
or intrinsic skyrmion Hall angle.
In the overdamped limit, the system
undergoes a transition from a disordered phase
to a moving smectic state in which
the velocity fluctuations are strongly anisotropic
and have the largest magnitude parallel to the direction of motion.
In the center of mass frame of the smectic state,
the particle displacements
are either diffusive or superdiffusive in the direction of motion but subdiffusive in the
perpendicular direction.
For a finite intrinsic skyrmion Hall angle,
the system is disordered at low drives, while at higher drives it
transitions into a moving
crystal state
which exhibits isotropic velocity fluctuations and has
subdiffusive particle displacements in the center of mass frame both parallel
and perpendicular to the direction of motion.
The isotropic nature of the
moving skyrmion crystal is a result of  the Magnus term,
which generates velocity fluctuations
perpendicular to the force fluctuations
experienced by the moving skyrmions due to the pinning sites.
In general, moving skyrmion lattices are more ordered than moving
vortex lattices and the dynamic shaking temperature
is isotropic for skyrmions but anisotropic in vortex systems.
We show that the
velocity noise power spectra can be used to
identify the  transition from the plastic flow or moving liquid
state to the moving crystal state in skyrmion systems
since it changes from a broad band noise signal
to a narrow band noise signal.
In the moving crystal state, velocity noise power peaks
appear at the washboard
frequency, which
permits the calculation of the velocity and lattice spacing of the moving skyrmion lattice.
We find that the moving skyrmion lattice can exhibit discrete switching
events associated with
global lattice reorientations due to the
dependence of the skyrmion Hall angle on the magnitude of the external drive.
These switching events produce changes in the noise fluctuations as well
as rotations in the structure factor.
For small but finite intrinsic skyrmion
Hall angles, the system exhibits a mixture of both vortex-like and skyrmion-like
dynamical behavior.   At lower drives the
system first dynamically orders into a moving smectic state,
but at higher drives there is a transition into a moving crystal which
is associated with both a change in the direction of skyrmion motion
as well as pronounced changes in the velocity noise signal.
Skyrmion velocity fluctuations could be measured by direct imaging of moving
skyrmions, measuring fluctuations in the topological Hall resistance,
performing magnetic noise measurements with Hall probes,
or measuring the time dependence of the structure factor.

\begin{acknowledgments}
  This work was carried out under the auspices of the
  NNSA of the U.S. DoE at LANL under Contract No.
  DE-AC52-06NA25396.
\end{acknowledgments}

\end{document}